\documentclass[twocolumn,aps]{revtex4}
\input epsf
\usepackage{graphicx}
\usepackage{dcolumn}
\usepackage{bm}
\begin{document}

\title{Phase transitions and crossovers in reaction-diffusion
models with catalyst deactivation}

\author{T.G. Mattos}
 \email{tgmattos@if.uff.br}
 \affiliation{Instituto de F\'{\i}sica, Universidade Federal Fluminense, 
Av. Litoranea s/n, Campus da Praia Vermelha, Niteroi RJ 24210- 340, Brazil.}

\author{F\'abio D. A. Aar\~ao Reis}
 \email{reis@if.uff.br}
 \altaffiliation[~~Also at ]{Department of Chemistry, University of Wisconsin - Madison, WI 53706, USA.}
 \affiliation{Instituto de F\'{\i}sica, Universidade Federal Fluminense, 
Av. Litoranea s/n, Campus da Praia Vermelha, Niteroi RJ 24210- 340, Brazil.}

\date{\today}% It is always \today, today,

\begin{abstract}
The activity of catalytic materials is reduced during operation 
by several mechanisms, one of them being
poisoning of catalytic sites by chemisorbed impurities or products.
Here we study the effects of poisoning in two reaction-diffusion models
in one-dimensional lattices with randomly distributed catalytic sites.
Unimolecular and bimolecular single-species reactions are considered, 
without reactant input during the operation.
The models show transitions between a phase with
continuous decay of reactant concentration and a phase with asymptotic 
non-zero reactant concentration and complete poisoning of the catalyst.
The transition boundary depends on the initial reactant and catalyst
concentrations and on the poisoning probability.
The critical system behaves as in the two-species annihilation reaction,
with reactant concentration decaying as $t^{-1/4}$ and the catalytic
sites playing the role of the second species.
In the unimolecular reaction, a significant crossover to the asymptotic 
scaling is observed even when one of those parameters is $10\%$ far from 
criticality. Consequently, an effective power-law decay of concentration
may persist up to long times and lead to an apparent change in the reaction
kinetics. In the bimolecular single-species reaction, the critical scaling 
is followed by a two-dimensional rapid decay, thus two crossovers are found.

\end{abstract}

\maketitle

\section{Introduction}

% small changes
Simple models for reactions of diffusing species have been intensively studied in
the last decades \cite{hba,marro,rice}, such as
trapping reactions, annihilation or coagulation 
of a single species and two-species annihilation. In low dimensions, 
these apparently simple models yield a wide range of nontrivial kinetic behaviors
because mean-field theories (laws of mass action) fail.
One-dimensional media may be a realistic description of the structure of catalysts 
with long and narrow pores, such as zeolites and porous oxides,
or may represent step edges of two-dimensional surfaces where reactants
preferrably adsorb. The former materials have inhomogeneous distributions of
catalytic centers where the reactions take place, which motivated recent
theoretical models. If the reaction is
unimolecular with volatile products, then that inhomogeneity can be represented 
by trapping models
\cite{dv,bunde1997,barkema}, with the traps playing the role of catalytic sites. 
For other reaction mechanisms, the disorder in the spatial distribution of the 
catalytic centers was also considered in recent works
\cite{oshanin1998,toxvaerd,mandache,mobilia,oshaninJPA}. In some cases, it
leads to nontrivial dependence of the reaction rates on diffusion 
coefficients and catalyst density. Effects of catalyst geometry and conditions to 
improve the efficiency of a reaction process were also discussed 
in recent works \cite{mcleod,cwiklik,spillover}.

% two paragraphs in one
A phenomenon of great economic impact on industrial processes is catalyst
deactivation, which is the reduction of catalytic activity due to blocking of active
sites, distortion or blockage of the porous structure, sintering of metal particles 
etc \cite{bartholomew,spencer,cerqueira}.
One of the possible deactivation mechanisms is poisoning. It occurs when a reactant,
product or impurity is strongly chemisorbed in the active sites, preventing
further reactions but not 
affecting the diffusion of reactants and products along the pores.
For instance, poisoning of metals by sulphur species
($H_2S$, $SO_2$, ${SO_4}^{2-}$ etc) is a
problem in many catalytic processes, such as hydrogenation, methanation, 
Fischer–Tropsch synthesis, steam reforming and fuel cell power production.
In other cases, large aggregates are formed inside the catalyst pores; this
is called coking or fouling and eventually leads to pore blockage.
Due to its technological relevance, deactivation mechanisms are frequently included 
in models of catalytic processes \cite{bartholomew,froment}.
The main interest of those models is to predict the time evolution of the
catalytic activity and the turnover frequency
with a continuous flux of reactants, thus they are usually designed 
for a particular application.

% changes in comments on chemical motivation and limitations of the model
However, there is no systematic study of the effects of
deactivation on the kinetics of simple reaction-diffusion models (e. g. trapping or 
annihilation). The aim of this work is to fill this gap by introducing one-dimensional
models for catalyzed reactions of diffusing reactants with poisoning of catalytic sites.
We will consider initial conditions with uniform distributions of reactants and no
external flux. These conditions are far from those encountered in real
catalytic processes in porous media, but they are interesting as a first step to understand 
possible changes in those systems kinetics.

We will study these models with scaling concepts supported by numerical simulations.
Unimolecular and bimolecular (same species) reactions occuring
upon contact of the reactants with the catalytic sites are considered.
Poisoning is represented by the permanent blockage of those sites, with a given
probability, immediately after the reaction takes place. 
Two limiting cases of this type of model anticipate the 
presence of kinetic transitions:
for very low deactivation rates and low initial reactant concentration, 
poisoning is negligible, thus
that concentration decays similarly to the case without deactivation; however,
for high deactivation rates and high initial reactant 
concentration, the catalyst will be rapidly poisoned and part of the reactants will
not be consumed. This analysis raises the question of how these systems behave when
deactivation rates, initial reactant concentrations and catalyst loadings change, 
and which types of transition occur.

We will show that two phases exist in those systems. In one phase,
the reactant concentration continuously decay to zero,
with the same asymptotic scaling of the main reaction. In the other phase,
there is a non-zero reactant concentration at long times.
In the boundary between those phases, the system behaves as in the 
two-species annihilation reaction,
with reactant concentration decaying as $t^{-1/4}$ in one dimension - the catalytic
sites play the role of the second species in this case. Near a transition point,
we will show a long crossover to the asymptotic scaling of the phase of
decaying reactant concentration. Thus, a correct interpretation
of the main reaction kinetics becomes difficult, particularly with short time 
simulation data. For bimolecular reactions, the deactivation
also enlarges the time interval of an intermediate two-dimensional scaling,
where the concentration decay is faster than both the short-time and the
long-time ones. These crossovers may also have important consequences in the
interpretation of experimental results.

The rest of this paper is organized as follows. In Sec. II we consider unimolecular
reactions with inert products taking place at random catalytic sites (traps) 
subject to deactivation. In Sec. III, bimolecular single-species reactions are
considered and in Sec. IV we summarize our results and present our conclusions.

\section{Trapping reactions with trap deactivation}

In this model, defined in a one-dimensional lattice,
we consider unimolecular reactions with volatile products,
which occur when the reactant is in contact with the catalyst.
Initially, catalytic sites (C) are randomly distributed
with coverage $\sigma_0$ and are immobile.
The other sites are labeled as non-catalytic or empty sites.
At $t=0$, a concentration $\theta_0$ of reactants (species A) is randomly distributed
through the non-catalytic sites - consequently, the condition $\sigma_0+\theta_0\leq 1$
is obeyed. For $t>0$, reactants A diffuse with coefficient $D$, i. e. 
each reactant executes an average
of $D$ random steps to neighboring sites per unit time (the lattice parameter is
the length unit), and they obey the excluded volume condition.

The unimolecular reaction corresponds to a trapping reaction \cite{hba} in the form
\begin{equation}
A+C \to C ,
\label{unimolecular}
\end{equation}
which immediately occurs when A occupies the same site of C.
With probability $p$, the catalytic site is deactivated, i. e. C is 
converted into an empty site (consequently, it does not affect diffusion 
of the other reactants). The reaction with subsequent poisoning corresponds to 
the reaction scheme
\begin{equation}
A+C \to 0 ,
\label{unimolecularp}
\end{equation}
where $0$ represents an empty site.
Thus, the model may be viewed as a competition between the reaction (\ref{unimolecular}),
with probability $1-p$, and the reaction (\ref{unimolecularp}), with probability $p$.
This is illustrated in Fig. 1a.

\begin{figure}[!h]
\begin{center}
\includegraphics[width=0.4\textwidth]{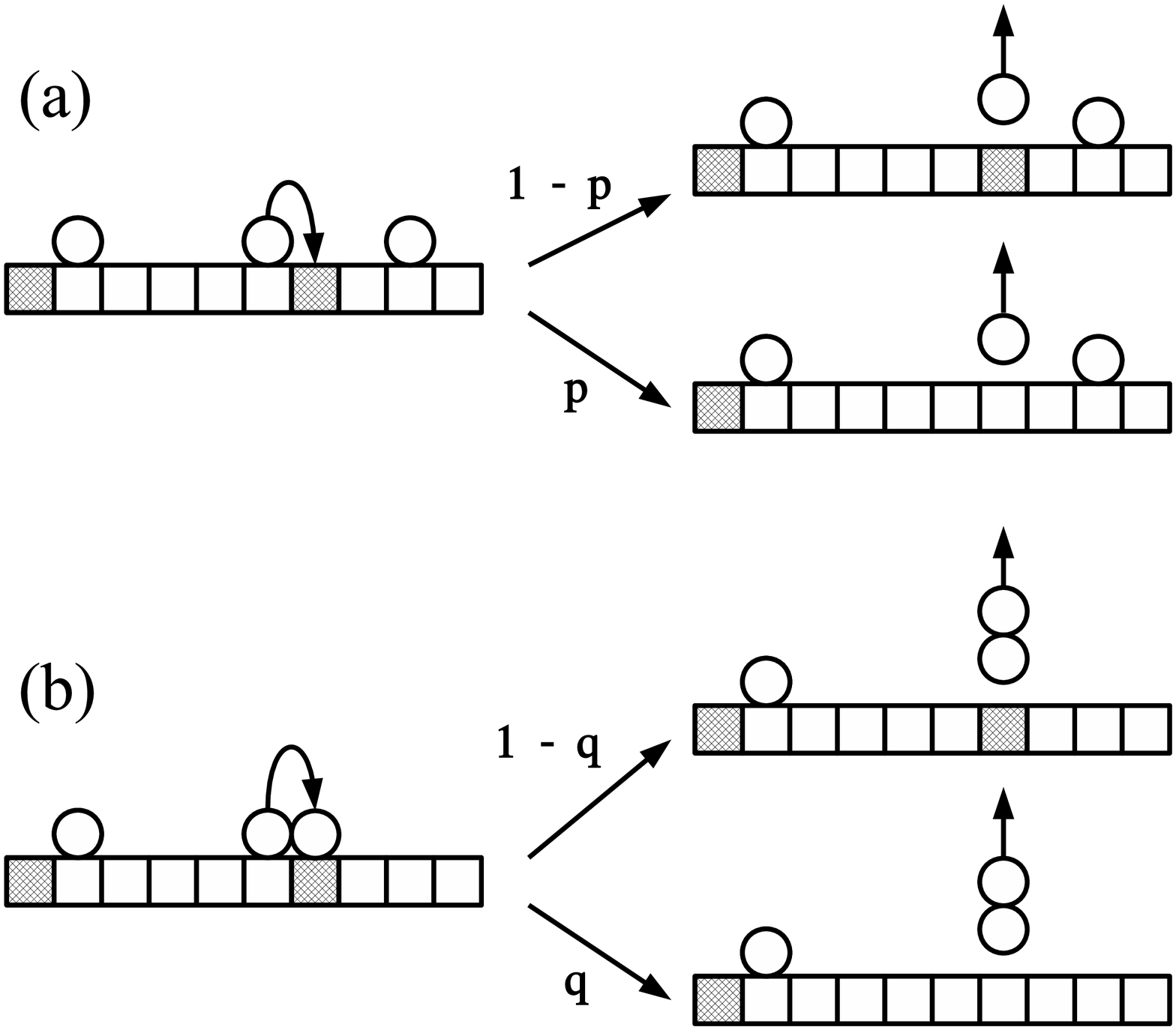}
\caption{Schemes of (a) unimolecular and (b) bimolecular single-species reactions in a line with catalytic (filled squares) and non-catalytic (open squares) sites. In both cases, the step of a reactant towards a catalytic site is shown in the left and the possible outcomes are shown in the right, with the associated probabilities.}
\label{fig1}
\end{center}
\end{figure}

As time evolves, the catalyst coverage and the reactant concentration are 
respectively denoted by $\sigma\left( t\right)$ and $\theta\left( t\right)$.

We performed simulations of this model in lattices of length $N=2^{20}$ (more than
one million sites), with several different values of the parameters $\sigma_0$, 
$\theta_0$ and $p$. For each set of parameters, typically ${10}^3$ different
realizations were averaged. Diffusion coefficients of species A were $D=1$ in
most cases. Maximum simulation times were up to ${10}^8$ units under these conditions.

\subsection{Long time scaling}

For $p=0$, we recover the well-known trapping problem with static traps. 
Since there is no deactivation,
all the reactants initially placed in a segment between two consecutive traps are
annihilated in one of the edge traps. For a fixed segment size $L$, the decay
is a simple exponential in $t^{1/2}/L$. However, the average over (Poisson distributed)
segment lengths is dominated by trapping in large segments \cite{hba}, which
leads to the Donsker-Varadan \cite{dv} result
\begin{equation}
\theta \sim \exp{\left[ -a\rho^{2/3}{\left( Dt\right)}^{1/3}\right]} ,
\label{expt13}
\end{equation}
where $\rho$ is the trap density and $a\equiv 3/2 {\left( 2\pi^2\right)}^{1/3}$. 
In the case $p=0$, we have $\rho=\sigma_0$. 

For $p$ small but non-zero, the concentration of reactants rapidly decays to very
low values, while only a fraction of the initial catalytic sites is deactivated.
The annihilation of a concentration
$\theta_0$ of species A leads to poisoning of a density $p\theta_0$ of catalytic sites.
Thus, at long times, the trap density is $\sigma = \sigma_0-p\theta_0$ and the
reactant concentration $\theta$ is much smaller. 
In this case, reaction (\ref{unimolecular}) is asymptotically dominant, thus
a decay as in Eq. (\ref{expt13}) is expected.

% small changes to highlight the unexpected good scaling for nonzero p
Simulations with small values of $p$ confirm this result, as shown in Fig. 2a.
They also show another interesting and somehow surprising feature:
while results for $p=0$ show significant corrections to the dominant scaling
(Eq. \ref{expt13}), the results for non-zero $p$ show that stretched exponential
decay with negligible corrections
(the data for $p=0.1$ in Fig. 2a fits a straight line in three time decades).
The corrections for $p=0$ were observed in previous simulation work \cite{bunde1997},
and for this reason it took many years for the original analytical
prediction in two and three dimensions \cite{dv} to be confirmed 
numerically \cite{barkema}.

\begin{figure}[!h]
\begin{center}
\includegraphics[width=0.4\textwidth]{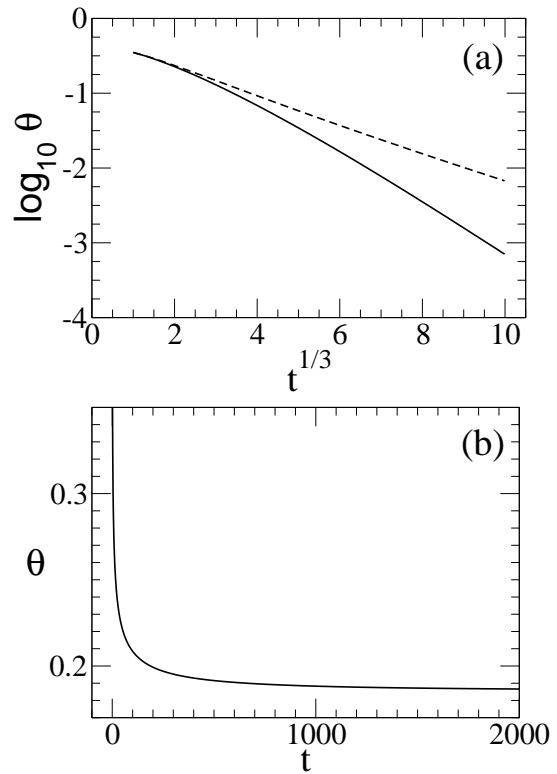}
\caption{Reactant concentration as a function of time for the unimolecular reaction with $\sigma_0 =0.15$, $\theta_0 =0.4$: (a) Exponential decay for $p=0$ (solid line) and $p=0.1$ (dashed line). (b) Decay towards a non-zero value for $p=0.7$.}
\label{fig2}
\end{center}
\end{figure}

From the above discussion, we expect the same asymptotic behavior for any initial
condition where $\sigma_0 > p\theta_0$. On the other hand, if $\sigma_0 < p\theta_0$,
a finite reactant concentration  $\theta = \theta_0 - \sigma_0 / p$ is found at 
long times, while the trap density continuously decays. This regime is dominated 
by reaction (\ref{unimolecularp}), with excess concentration of A relatively to C.
Fig. 2b confirms this result for $\theta_0=0.4$, $\sigma_0=0.15$ and $p=0.7$, which
gives a remaining density of reactants $\theta\approx 0.186$.

A transition is found when
\begin{equation}
\sigma_0 = p\theta_0 ,
\label{transition1}
\end{equation}
which corresponds to a surface in the $\left( \sigma_0,\theta_0,p\right)$ space.
In this case, the reactant concentration and the catalyst coverage decrease by
the same factors at all times.

% small changes
The continuous increase of the distance between consecutive 
catalytic sites certainly rules out the arguments that lead to the result in Eq. 
(\ref{expt13}). Instead, in a transition point 
this system resembles the two species annihilation reaction of Eq. 
(\ref{unimolecularp}) with balanced concentrations of species A and C.
According to previous work \cite{wilczek,hba}, in the reaction (\ref{unimolecularp})
with equal initial concentrations $\rho_0$ of A and C in one dimension, 
the reactant concentration $\rho$ decays as
\cite{wilczek,hba}
\begin{equation}
\rho \sim \frac{{\rho_0}^{1/2}}{{\left[\left( D_A+D_C\right)t\right]}^{1/4}} ,
\label{twospeciesdecay}
\end{equation}
where $D_A$ and $D_C$ are diffusion coefficients.
Note that this scaling law does not change if one species
is static, which is the case of the catalyst C in our model.

% new paragraph, comments on critical decay of reactants
The decay as $t^{-1/4}$ is confirmed for several transition points, as shown
in Fig. 3. However, the dependence with the initial reactant concentration is
different from the square-root law of Eq. (\ref{twospeciesdecay}):
for fixed $\sigma_0$, $\theta$ is proportional to $\theta_0$.
This occurs because the kinetics leading to Eq. (\ref{twospeciesdecay}) is
a balanced annihilation of AC pairs, but in our model the annihilation of an
AC pair requires trapping of $1/p$ reactants (A) in the average.
Thus, $\rho_0$ must be replaced by $\theta_0/p = {\theta_0}^2/\sigma_0$ in
Eq. (\ref{twospeciesdecay}).

\begin{figure}[!h]
\begin{center}
\includegraphics[width=0.4\textwidth]{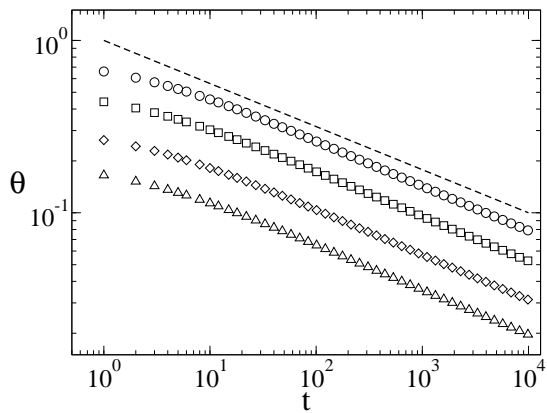}
\caption{Critical decay of reactant concentration in the unimolecular reaction model with $\sigma_0 = 0.15$: $\theta_0 =0.75$, $p =0.2$ (circles); $\theta_0 =0.5$, $p =0.3$ (squares); $\theta_0 =0.3$, $p =0.5$ (diamonds); $\theta_0 =0.1875$, $p =0.8$ (triangles). The solid line shows a decay as $t^{-1/4}$.}
\label{fig3}
\end{center}
\end{figure}

\subsection{Crossover scaling} 

Here we focus on the case where the system is near the critical
surface (Eq. \ref{transition1}) but $p<p_c\equiv \sigma_0 /\theta_0$, i. e.
there is a continuous decay of the reactant concentration. At short times,
this decay follows the critical power law of 
Eq. (\ref{twospeciesdecay}), but it crosses over to the stretched exponential 
behavior (Eq. \ref{expt13}) at long times. This is illustrated in Figs. 4a and 4b 
for fixed values of $\sigma_0$ and $\theta_0$ and various $p$.

\begin{figure}[!h]
\begin{center}
\includegraphics[width=0.4\textwidth]{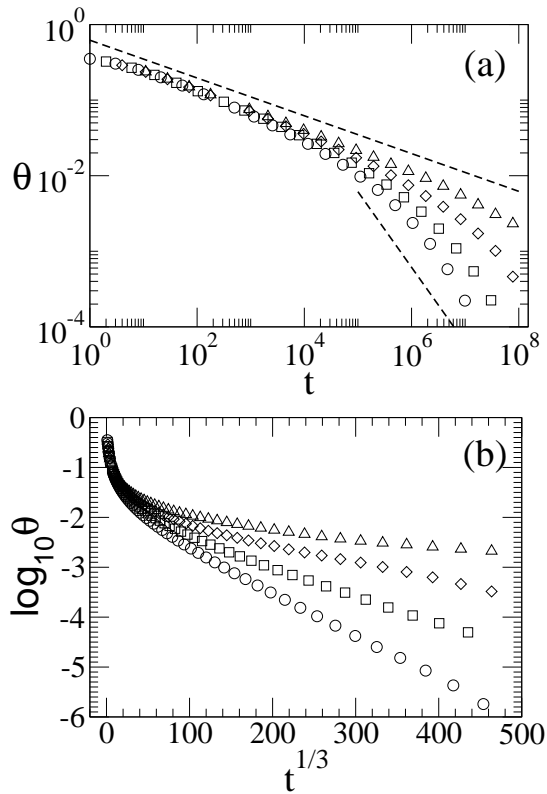}
\caption{(a) Short time and (b) long time decays of reactant concentration in the unimolecular reaction model near the critical boundary with $\sigma_0 =0.15$, $\theta_0 =0.4$: $p=0.34$ (circles), $p=0.35$ (squares), $p=0.36$ (diamonds), $p=0.37$ (triangles); critical point is at $p=0.375$. In (a), the dashed lines have slopes $-1/4$ and $-1$.}
\label{fig4}
\end{center}
\end{figure}

% small changes, new comments
One interesting point is that, even with one parameter being $10\%$ distant from the
critical value (e. g. $p=0.34$ in Fig. 4a), the concentration decay for
$\theta >{10}^{-4}$ still resembles a power law. The exponent of this apparent
power law depends on the chosen time range, varying from $-1/4$ for $1\leq t\leq {10}^3$
to almost $-1$ for ${10}^6\leq t\leq {10}^7$, as illustrated in Fig. 4a.
Closer to $t_c$, apparent power laws
are found for longer times. This is a surprisingly long crossover
due to the remarkable difference between the scaling of Eq. (\ref{twospeciesdecay})
and the asymptotic one of Eq. (\ref{expt13}).

% small change
At long times, the remaining density of catalytic sites is proportional to $p_c-p$.
Eq. (\ref{expt13}) gives a reactant concentration
\begin{equation}
\theta \sim \exp{\left[ -b{\left( Dt\right)}^{1/3}\right]} , 
b\sim {\left( p_c-p\right)}^{2/3} .
\label{crossover1}
\end{equation}
However, the scaling of $\theta$ on $p_c-p$ is difficult to be confirmed
numerically (e. g. in Fig. 4b) because the data close to $p_c$
shows huge corrections to the (noncritical) asymptotic decay of Eq. \ref{expt13}.

From the above result, we can estimate the time in which the apparent power 
law decay  crosses over to the exponential one. Matching the time scaling in
Eqs. (\ref{crossover1}) and (\ref{twospeciesdecay}) (the latter with no amplitude
depending on $p_c-p$), we obtain a crossover time of order
\begin{equation}
t_c \sim {\left( p_c-p\right)}^{-2} {\left[ |\ln{\left( p_c-p\right)}|\right]}^3 .
\label{crossovertime}
\end{equation}
This helps to understand the long crossover in this system: for the values
of $p$ shown in Figs. 4a and 4b, we expect $t_c$ to range approximately 
from ${10}^4$ to ${10}^7$. 
Experimentally, it has an important consequence: the reaction
kinetics may be incorrectly identified if one observes the time evolution of the 
concentration decay (or the turnover rate) in a restricted time range and disregards
the effects of deactivation.

\section{Annihilation reactions with trap deactivation}

The second reaction mechanism studied in this work is bimolecular with a
single species. The model corresponds to annihilation reactions of this species
limited to catalytic sites of a lattice.

Again we consider initial random distributions of static catalytic sites C, with
coverage $\sigma_0$, and diffusing reactants A with initial concentration $\theta_0$.
The bimolecular reactions between two A which meet at a catalytic site is 
represented by
\begin{equation}
A+A+C \to C ,
\label{bimolecular}
\end{equation}
It corresponds to the well known one-species annihilation model ($A+A\to 0$) restricted 
to the set of catalytic sites, as in Ref. \cite{oshanin1998}.
The poisoning is also represented by the permanent blockage of a catalytic site after
a reaction, and occurs with probability $q$. It corresponds to the reaction
\begin{equation}
A+A+C \to 0 .
\label{bimolecularq}
\end{equation}
The competitive model is illustrated in Fig. 1b.

\subsection{Long time scaling}

When $q=0$, we have the model proposed by Oshanin and Blumen \cite{oshanin1998}, 
in which the density of reactants decays as
\begin{equation}
\theta \sim 1/{\left( Dt\right)}^{-1/2}
\label{t12}
\end{equation}
in one dimension. This dominant contribution to the decay does not involve the 
catalyst coverage, which can be explained as follows. The characteristic time for
two  consecutive reactants 
to meet at the same site is $\tau_{AA}\sim 1/\left( D\theta^2\right)$. After they
meet, the closest trap is inside a neighborhood of size $1/\sigma_0$. Thus,
the typical time for those reactants to meet at that trap is
$\tau_{AAC}\sim 1/\left( D{\sigma_0}^2\right)$. At long times, $\theta\ll\sigma_0$,
thus we have $\tau_{AAC}\ll\tau_{AA}$ and the long-time scaling is dominated 
by $\tau_{AA}$, which leads to Eq. (\ref{t12}).
However, the catalyst coverage plays an important role at intermediate
times, as will be shown in Sec. \ref{crossoverscalingAA}.

For small $q$, we expect the same scaling as in Eq. (\ref{t12}).
At long times, $\theta\ll\theta_0$, thus the density of annihilated A particles 
tends to $\theta_0$ and a density
$q\theta_0 /2$ of catalytic sites will be deactivated. Thus, the final catalytic
coverage is $\sigma_0-q\theta_0 /2$. Simulation
results shown in Fig. 5 confirm these predictions.

\begin{figure}[!h]
\begin{center}
\includegraphics[width=0.4\textwidth]{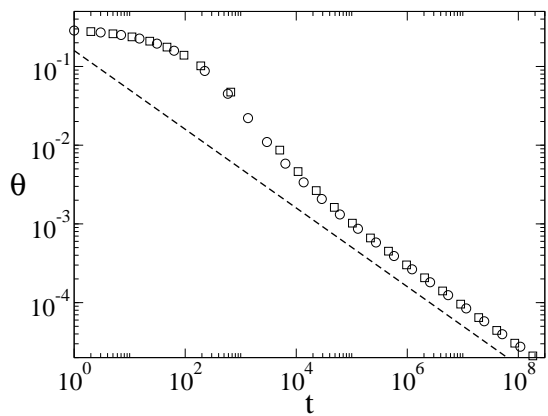}
\caption{Decay of reactant concentration far from the critical boundary for the bimolecular single-species model with $\sigma_0 =0.1$, $\theta_0 =0.3$: $q=0$ (circles) and $q=0.1$ (squares). The dashed line shows a decay as $t^{-1/2}$.}
\label{fig5}
\end{center}
\end{figure}

We expect a transition between a phase of decaying reactant
concentration and a phase of non-zero long time concentration when
\begin{equation}
\sigma_0 = q\theta_0 /2 .
\label{transition2}
\end{equation}
In this critical surface, reactant concentration and catalyst coverage decrease by 
the same factors as time increases.

The three-particle reaction (\ref{bimolecularq}) and an analogy with the case of Sec. II 
could suggest that the scaling of the concentration at the transition point is that
of a three-particle annihilation model. Since the
catalyst coverage does not play a role in the decay for $q=0$ (Eq. \ref{t12}),
we might think that the time $\tau_{AA}$ governs the scaling at the transition.
However, this is not the truth. Similarly to the 
unimolecular case of Sec. II, there is accummulation of C at some
regions where the initial concentration of A was locally small. Also, there is
accummulation of A at neighboring regions, where its initial concentration
was locally high. Thus, the coarsening is again dominated by
the annihilation time of neighboring domains, which scale as
$1/\theta^4$ \cite{hba,wilczek}, and not 
by the time $\tau_{AA}$. Thus, this transition also scales as in two-species 
annihilation model, although three particles are involved in the reactions
(\ref{bimolecular} and \ref{bimolecularq}).

These conclusions are supported by simulation results. In Fig. 6a, we show 
the average number of consecutive A, $\langle S_A\rangle$, and of consecutive C, 
$\langle S_C\rangle$, as function of time, in one point of the critical
surface. They scale approximately as
\begin{equation}
\langle S_A\rangle \sim \langle S_C\rangle \sim t^{1/4} .
\label{SASC}
\end{equation}
Fig. 6b shows the decay of the reactant concentration approximately as $t^{-1/4}$ (Eq. 
\ref{twospeciesdecay}) in three points of the critical surface.
All these results are typical of the two-species annihilation
problem \cite{hba}. Note that the total size of domain A (C) includes the 
reactants (catalysts) and the empty sites between them, thus this size
increases as $t^{1/2}$ (in the literature, this total size is the one usually
referred to as the domain size).

\begin{figure}[!h]
\begin{center}
\includegraphics[width=0.4\textwidth]{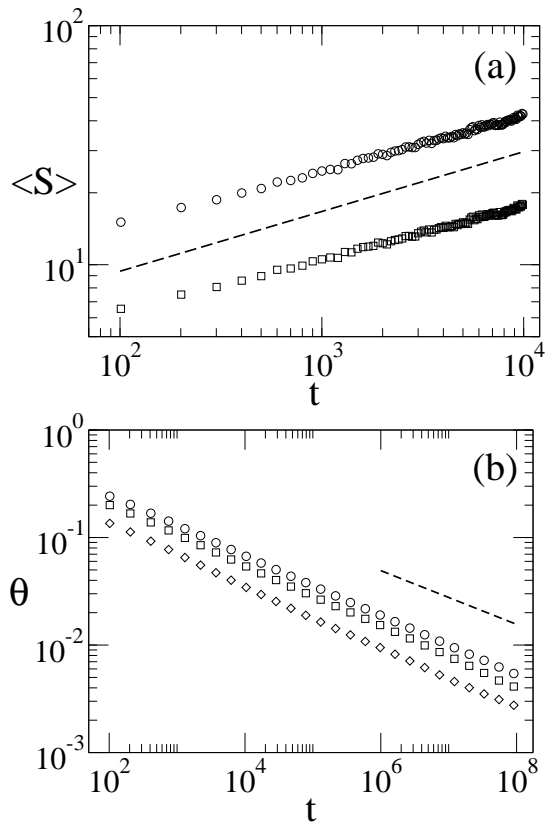}
\caption{(a) Average number of consecutive reactants (circles) and consecutive catalytic sites (squares) in a critical point of the bimolecular single-species reaction model: $\sigma_0 =0.4$, $\theta_0 =10/11$, $q=0.88$. The dashed line shows a scaling as $t^{1/4}$. (b) Critical decay of the reactant concentration in the same model with $\sigma_0=0.1$: $q=0.4$, $\theta_0=0.5$ (circles); $q=0.5$, $\theta_0=0.4$ (squares);  $q=0.8$, $\theta_0=0.25$ (diamonds). The dashed line shows a decay as $t^{-1/4}$.}
\label{fig6}
\end{center}
\end{figure}

% comments on initial concentration dependence
In Fig. 6b, we observe that the critical $\theta$ depends on the
initial concentration in the same way as the unimolecular reaction, i. e., with
$\rho$ in Eq. (\ref{twospeciesdecay}) replaced by 
$\theta_0/q\sim {\theta_0}^2/\sigma_0$.

For $\sigma_0 <q\theta_0 /2$, the catalyst coverage vanishes asymptotically,
while a finite number of reactants is present. The final (non-consumed) reactant 
concentration
is $\theta_0-2\sigma_0/q$. These results were also confirmed by simulation.

\subsection{Crossover scaling}
\label{crossoverscalingAA}

The above analysis suggests that a crossover between two power law decays 
($t^{-1/4}$ and $t^{-1/2}$) would be observed near a transition
point. However, again the situation is much more complex. Ref.
\protect\cite{oshanin1998} analyzed the problem with $q=0$ (no deactivation)
and showed that, for small catalyst concentration, there
is a long time interval with effective two-dimensional decay of the
concentration:
\begin{equation}
\theta \sim \ln{t}/t .
\label{decayAAinterm}
\end{equation}

Small catalyst concentration is actually the case as we approach the critical 
point. Consequently, not only two but three scaling regions are expected:
for short times, the critical decay
as $t^{-1/4}$, while the orders of magnitude of
$\sigma$ and $q\theta /2$ are not very different (see Eq. \ref{transition2});
for intermediate times, the much faster power law decay of Eq. 
\ref{decayAAinterm}; for $t\gg 1/{\sigma_\infty}^2$
\cite{oshanin1998}, where $\sigma_\infty$ is the asymptotic concentration of the
catalyst, the asymptotic scaling (\ref{t12}) is found. Since
$1/{\sigma_\infty}^2 \sim {\left(q_c-q\right)}^{-2}$ near criticality, 
the deactivation significantly enlarges the time window of the
intermediate scaling.

These results are illustrated in Fig. 7 for several values of $q$. For $q=0.21$,
($16\%$ far from criticality), a very long intermediate regime is observed,
and the asymptotic decay of Eq. (\ref{t12}) is found only for $t\gg {10}^6$ 
and concentrations below ${10}^{-4}$. Note that this decay is not dependent
on initial concentrations, as shown by the data for $q=0.21$ and $q=0.22$.
Closer to the critical point, the asymptotic decay is not 
observed even with concentrations below ${10}^{-4}$. Compared to the 
unimolecular reactions (Sec. II), finding the true reaction kinetics from reactant
or product concentration data is much trickier in this case due to the presence
of two crossovers.

\begin{figure}[!h]
\begin{center}
\includegraphics[width=0.4\textwidth]{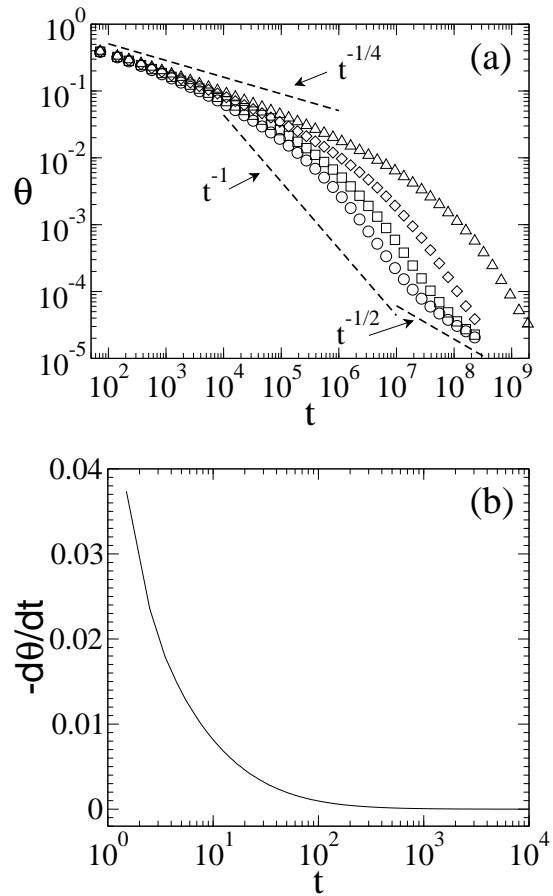}
\caption{Reactant concentration near the critical point of the bimolecular single-species reaction model with $\sigma_0 =0.1$, $\theta_0 =0.8$: $q=0.21$ (circles),  $q=0.22$ (squares), $q=0.23$ (diamonds), $q=0.24$ (triangles); critical point is at $q=0.25$. Dashed lines indicate three types of power-law decay present in this system.}
\label{fig7}
\end{center}
\end{figure}

\section{Discussion and conclusion}

We studied reaction-diffusion models in heterogeneous one-dimensional
lattices with catalytic and non-catalytic sites and possible deactivation
of catalytic sites by poisoning. These models show a transition 
between a phase with
continuous decay of reactant concentration and survival of catalytic
sites, and a phase with asymptotic non-zero reactant concentration and
complete poisoning of the catalyst. The critical systems behave as
the two-species annihilation model with stoichiometric concentrations of
reactants. For each reaction mechanism, the
transition boundary depends on the initial concentrations of reactant 
and catalyst and the probability of deactivation.
We analyzed models of unimolecular and bimolecular single-species
reactions, and showed that, in a transition point, the reactant concentration
decays as in the two-species annihilation reaction. In the decaying phase but
near a critical point of the unimolecular reaction, a long crossover to the
asymptotic scaling is found. Thus, short time simulations show effective 
power-law decays instead of the stretched exponential, which suggests a
change in the reaction kinetics. The situation is much more complex with the
bimolecular single-species reaction because the small catalyst density near
a critical point leads to an apparent two-dimensional decay of reactant 
concentration.
Thus, two crossovers are observed and the asymptotic decay is found only at
extremely long times, even with one parameter being more than $10\%$ far from
criticality. 

Our results resemble transitions in other competitive reaction-diffusion models
which were not explicitly interpreted as effects of catalyst
deactivation. Sanchez and co-workers studied the case where reaction 
$A+B\to B$ competes with $B+C\to C$ (double trapping) or $B+C\to 0$ (trapping
plus annihilation) \cite{sanchez1,sanchez2}, showing that
the initial concentrations and the reaction rates also determine the 
asymptotic behavior. Further work on these and related
reactions also considered anisotropic or anomalous diffusion of reactants, 
different forms of reactant input, and power-law decays of reaction rates, 
which lead to a wide range of interesting kinetic behaviors
\cite{bray,lee,merkin,rasaiah,turban,yuste}. For instance, the model of 
Yuste et al \cite{yuste}
for trapping by subdiffusive traps with vanishing density shows a transition 
with scaling similar to our unimolecular model.

Despite the simplicity of the reaction-diffusion models presented here,
experimental 
applications are possible. For instance, the depletion zones of a trapping 
model with a single trap were already observed in photobleaching of fluorescein
dye by a focused laser beam \cite{trap1,trap2}, and models with competitive
reactions were already used to represent patterns of a reaction between
${Cr}^{3+}$ and Xylenol Orange \cite{hecht}. Thus, we believe
that our work may also have applications to systems with catalyzed reactions
and deactivation by poisoning, particularly to help the interpretation of
kinetic data.

\vskip 0.5cm

{\par\noindent\bf Acknowledgments.} We acknowledge support from CNPq and
Faperj (Brazilian agencies) to our simulation laboratory at UFF, Brazil.
TGM acknowledges a grant from CAPES (Brazil) and FDAAR acknowledges support from CNPq for his visit to UW - Madison.

\end{document}